\documentclass[12pt]{article}
\input amssym.def
\input amssym
\newcommand{\be}{\begin{equation}}
\newcommand{\ee}{\end{equation}}

\title{Relativistic generalization of the inertial and
gravitational masses equivalence principle}
\author{Nikolai V. Mitskievich\thanks{Physics Department,
CUCEI, University of Guadalajara, Guadalajara, Jalisco, Mexico.}
\thanks{Postal address: Apartado Postal 1-2011, C.P. 44100,
Guadalajara, Jalisco, M\'exico. E-mail:
mitskievich03@yahoo.com.mx}}
\date{~}

\begin{document}

\maketitle

\begin{abstract}
The Newtonian approximation for the gravitational field equation
should not necessarily involve admission of non-relativistic
properties of the source terms in Einstein's equations: it is
sufficient to merely consider the weak-field condition for
gravitational field. When a source has electromagnetic nature, one
simply {\em cannot} ignore its intrinsically relativistic
properties, since there cannot be invented any non-relativistic
approximation which would describe electromagnetic stress-energy
tensor adequately, even at large distances where the fields become
naturally weak. But the test particle on which gravitational field
is acting, should be treated as non-relativistic (this premise is
required for introduction of the Newtonian potential $\Phi_{\rm
N}$ from the geodesic equation).
\end{abstract}

We use here (in parentheses if in a tetrad basis) Greek indices as
4-dimensional and Latin as 3-dimensional, $\varkappa=8\pi G$ ($G$
is the Newtinian gravitational constant), $R_{\mu\nu}=
{R^\alpha}_{\mu\nu\alpha}$, and spacetime signature as $+,-,-,-$.
Einstein's equations then read as $R^{(\mu)}_{(\nu)}-\frac{1}{2}R
\delta^{\mu}_{\nu}=-\varkappa T^{(\mu)}_{(\nu)}$, thus $R
=\varkappa T$, and $R^{(\mu)}_{(\nu)}=-\varkappa\left(
T^{(\mu)}_{(\nu)}-\frac{1}{2}T\delta^{(\mu)}_{(\nu)}\right)$. We
shall need only $00$-component of Einstein's equations, \be
\label{2.1.3} R^{(0)}_{(0)}=-\frac{\varkappa}{2}\left(T^{(0)}_{
(0)}-T^{(i)}_{(i)}\right). \ee

We call a source with $T^{(\nu)}_{(\nu)}=0$ intrinsically
relativistic since the spatial part of its stress-energy tensor is
of the same order of magnitude as the temporal component (cf. the
concept of a zero rest mass particle). An example is the Maxwell
electromagnetic field which has this property even of its static
solutions when any kind of motion is excluded. Similarly, a
perfect fluid with its energy-momentum tensor \be \label{TmunuFl}
T^{\rm pf}=(\mu+p)u\otimes u-pg \ee possesses this property in the
particular case of incoherent radiation ($\mu=3p$), and the tensor
(\ref{TmunuFl}) is written in the rest reference frame of the
fluid. There is also the case of stiff matter ($p=\mu$) in which
the sound propagates with the velocity of light; we say that such
objects are hyper-relativistic. Thus in the non-relativistic case
$\left(\left|T^{(i)}_{(i)}\right|\ll T^{(0)}_{(0)}\right)$ the
00-component of Einstein's equations reads \be \label{nonrel}
R^{(0)}_{(0)}\approx-\frac{\varkappa}{2}{T_{\rm
non-rel}}^{(0)}_{(0) }, \ee then in the intrinsically relativistic
case, \be \label{intrrel} R^{(0)}_{(0)} =-\varkappa{T_{\rm
intr.rel}}^{(0)}_{(0)}, \ee and finally in the hyper-relativistic
case, \be \label{hyprel} ~~\,~\,~R^{(0)}_{(0)}=-2\varkappa {T_{\rm
hyper-rel}}^{(0)}_{(0)}. \ee

The Newtonian approximation is found from the geodesic motion of a
non-relativistic test particle. Thus let us consider a static
spacetime with $g_{00}=1+2\Phi_{\rm N}$, $|\Phi_{\rm N}|\ll 1$ and
choose a 1-form basis as \be \label{2.1.5} \theta^{(0)}={\rm
e}^\alpha dt, ~ ~ \theta^{(k)}={g^{(k)}}_jdx^j. \ee Taking the
inverse triad, so that $dx^j={g_{(k)}}^j\theta^{(k)}$, $dt={\rm
e}^{-\alpha} \theta^{(0)}$, we find the necessary components of
1-form connections
${\omega^{(0)}}_{(l)}\equiv{\omega^{(l)}}_{(0)}=
\alpha_{,j}{g_{(l)}}^j\theta^{(0)}$, and finally from Cartan's
second structural equations, \be \label{R00} R^{(0)}_{(0)}=
g^{(l)(k)}{R^{(0)}}_{(l)(k)(0)}\approx {\rm e}^{-\alpha}\left({\rm
e}^\alpha\right)_{,i,j} g^{ij} \ee where $g^{ij}=-\delta^i_j ~ +$
higher-order terms (to be neglected). Since ${\rm e}^\alpha\approx
1+\Phi_{\rm N}$, $R^{(0)}_{(0)}\approx-\Delta\Phi_{\rm N}$
($\Delta$ is the usual Laplacian). Thus the Newton--Poisson
equations corresponding to (\ref{nonrel}), (\ref{intrrel}), and
(\ref{hyprel}), are \begin{eqnarray} \label{Nnonrel}
\textnormal{non-relativistic~~~}
\Delta\Phi_{\rm N}= 4\pi G\mu,\phantom{aaa,a}\\ 
\label{Nintrrel} \textnormal{intrinsically relativistic~~~}
\Delta\Phi_{\rm N}=8\pi G\mu\mbox{ ~ ~and}\\
\label{Nhyprel} \textnormal{hyper-relativistic~~~} \Delta\Phi_{\rm
N}=16\pi G\mu,\phantom{aa,a}
\end{eqnarray}
respectively (we wrote here the inertial mass density $\mu$ of the
source instead of $T^{(0)}_{(0)}$). For any perfect fluid the
Newton--Poisson equation takes the form \be \label{2.1.9}
\Delta\Phi_{\rm N}=4\pi G(\mu+3p), \ee so that for incoherent dust
the old traditional equation follows, but if the fluid represents
an incoherent radiation ($p=\mu/3$), the source term doubles (as
this is the case for electromagnetic source), and for the stiff
matter ($p=\mu$), it quadruples.

Since the equations (\ref{intrrel}) and (\ref{hyprel}) are exact
ones, they strictly express the equivalence principle already
generalized (to use an expression similar to ``already unified''
of J.A.~Wheeler) in standard general relativity. The conclusions
we came upon in this talk automatically add on relativistic
features to the principle traditionally formulated in standard
textbooks on general relativity as a completely non-relativistic
approximation (for both test particle and sources of Einstein's
equations) just as it was used by Einstein in his first attempts
to generalize the special relativity. But the Newtonian-type
potential is generated by a wide class of distributions of matter,
including intrinsically relativistic and hyper-relativistic cases:
the only restriction here consists of weakness of the field and
not the ``state of motion'' of the sources in Einstein's equations
(especially such an intrinsic property as to be relativistic which
is so often realized by static configurations when the very idea
of motion is out of question). Clearly, here we haven't used any
hypotheses at all.

As to the applications of this generalized principle of
equivalence, it is worth pointing out the (post-) post-Newtonian
approximations. Since some conclusions about validity of the
principle of equivalence come from observations of stellar
systems, a mere presence in them of intrinsically relativistic
distributed or localized objects (say, high density of any kind of
radiation, strong or widely distributed magnetic fields, existence
of stiff matter in cores of exotic stars, jets of
ultrarelativistic particles) would radically change interpretation
of the observational data if their proper understanding depends on
adequate description of the sources of gravitational field,
without any disregard for the pressure and stresses. These
conclusions should definitively lead to a revision of the old
problem of stability of young globular star clusters via the
virial theorem (when the electromagnetic radiation between the
stars is very intense) which seems to be done through approximated
methods only. This is also the central point of evolution of the
gravitation theory from Soldner \cite{Soldner} and Einstein-1911
\cite{Einst11} to Einstein-1915 \cite{Einst15}, resulted in
doubling [cf. (\ref{Nnonrel}) and (\ref{Nintrrel})] of the light
beams bending in the final self-consistent version of the theory.
This doubling has two sides: one is mentioned just above, and
another pertains to light beams and jets of ultra-relativistic
particles via the 3rd Newtonian law, see comments on both in Refs.
4, 5 and 7. Another problem is connected to the interesting and
stimulating question by D.~Brill, the Chairman of the parallel
Session GT4 at which this talk was delivered: How to relate
Einstein's first tentative considerations of photons' absorbtion
by a material sample, leading to its temperature rise, and the
corresponding increase of its masses, both inertial and
gravitating ones? My answer was that the gravitational mass does
not satisfy a conservation law, at least that which follows from
the Noether theorem \cite{Mits58,Mits06} under the general
relativistic invariance of the action integral, in a contrast to
the inertial mass, and it is clear that both masses cannot
simultaneously be conserved, {\it e.g.} in the process of light
absorbtion.

Finally, it should be emphasized once more that in this talk we
made a revision of a too long persistent old viewpoint, but not of
the sane and mature theory.

\end{document}